\documentclass[showpacs,preprint,prb]{revtex4}
\usepackage{amsmath}
\usepackage{epsfig}
\usepackage{graphicx}
\usepackage{dcolumn}
\usepackage{bm}

\newcommand{\w}{\omega}

\renewcommand{\b}{\beta}
\renewcommand{\a}{\alpha}

\renewcommand{\k}{{\bf k}}

\def\gl{\lower.35em\hbox{$\stackrel{\textstyle>}{\textstyle<}$}}

\def\gapp{\lower.35em\hbox{$\stackrel{\textstyle>}{\sim}$}}
\def\lapp{\lower.35em\hbox{$\stackrel{\textstyle<}{\sim}$}}

\begin{document}

\title{Dynamical polarization of graphene at finite doping}
\author{B. Wunsch$^{1,2}$, T. Stauber$^2$, F. Sols$^1$, and F. Guinea$^2$}
\affiliation{$^1$ Departamento de F\'{\i}sica de Materiales,
Facultad de Ciencias F\'{\i}sicas,
Universidad Complutense de Madrid, E-28040 Madrid, Spain.\\
$^2$ Instituto de Ciencia de Materiales de Madrid, CSIC, Cantoblanco,
E-28049 Madrid, Spain. }
\date{\today}

\begin{abstract}
The polarization of graphene is calculated exactly within the random
phase approximation for arbitrary frequency, wave vector, and
doping. At finite doping, the static susceptibility saturates to a
constant value for low momenta. At $q=2 k_{F}$ it has a
discontinuity only in the second derivative. In the presence of a
charged impurity this results in Friedel oscillations which decay
with the same power law as the Thomas Fermi contribution, the latter
being always dominant. The spin density oscillations in the presence
of a magnetic impurity are also calculated. The dynamical
polarization for low $q$ and arbitrary $\omega $ is employed to
calculate the dispersion relation and the decay rate of plasmons and
acoustic phonons as a function of doping. The low screening of
graphene, combined with the absence of a gap, leads to a significant
stiffening of the longitudinal acoustic lattice vibrations.
\end{abstract}

 \maketitle


%







\section{Introduction}

Recent progress in the isolation of single graphene layers has
permitted the realization of transport and Raman
experiments\cite{Exp} which have stimulated an intense theoretical
research on the properties of a monoatomic graphene sheet. Most of
the unusual electronic properties can be understood in terms of a
simple tight-binding approach for the $\pi $-electrons of carbon,
which yields a gapless linear band structure with a vanishing
density of states at zero doping.\cite{Wallace,McClure}

The electronic band structure of an undoped single graphene sheet
allows for a description of the electronic properties in terms of an
effective field theory which is equivalent to Quantum
Electrodynamics (QED) in 2+1 dimensions. Within such a framework
important physical quantities such as the electron self-energy or
the charge and spin susceptibilities can be calculated exploiting
the special symmetries of the model.\cite{Gon94}

Finite doping away from half-filling qualitatively changes the above
description, since it breaks electron-hole symmetry  and also the
pseudo Lorentz invariance needed for the equivalence with QED in two
space dimensions. One is thus forced to retreat to conventional
condensed-matter techniques such as e.g. Matsubara Green's
functions. This was recently done for the static susceptibility of graphene at finite doping~\cite{Ando06}.
Similar calculations for bulk graphite had been performed by Shung
\textit{et al.},\cite{Shung} who investigated the dielectric
function and the plasmon behavior. Studies of static screening in
graphene
based on the Thomas-Fermi approximation had also been considered.\cite%
{DiVincenzo,Katsnelson1} The relation between the polarization and
the transport properties of graphene has been recently discussed in
Refs.~\onlinecite{Ando06,Nomura1,Nomura2}. Thermo-plasma polaritons in graphene are discussed in cite{Vafek}.

In this paper, we calculate the dynamical polarization within the
random phase approximation (RPA) for arbitrary wavevector,
frequency, and doping. We discuss the method of calculation in the
next section. We present in Section III results for static
screening, where we analyze the Friedel oscillations induced by a
charged or magnetic impurity, comparing our results with those of
the two-dimensional electron gas (2DEG). Section IV discusses the
plasmon dispersion relation and lifetime. In Section V we analyze
the screening of the longitudinal acoustic modes by the conduction
electrons. Finally, Section VI presents the main conclusions of our
work. Some of the more technical aspects of the formalism are
explained in the Appendix.

\section{RPA calculation}

Within the effective mass approximation, and focusing on one of the
two unequal $K$-points, the Hamiltonian of an hexagonal
graphene sheet is given in the Bloch spinor representation
as\cite{Wallace}
\begin{align}
H=\hbar v_F \sum_\k H_\k\quad,\quad H_\k=\psi_\k^\dagger%
\begin{pmatrix}
-k_F & \phi_\k \\
\phi_\k^* & -k_F%
\end{pmatrix}
\psi_\k\;.
\end{align}
Here $\phi_\k=k_x+ik_y$ and $\psi_\k=(a_\k,b_\k)^T$, where $a_\k$ and $b_\k$
are the destruction operators of the Bloch states of the two triangular
sublattices. We have also introduced the Fermi wavevector $k_F$ which
is related to the chemical potential $\mu$ via $k_F=\mu/\hbar v_F$. The Fermi
velocity $v_F=3 a t/2 \hbar$ is determined by the carbon-carbon distance $%
a=1.42$ \AA \, and the nearest neighbor hopping energy $t=2.7$ eV
resulting in $v_F\simeq 9\times10^{5}$ m/s.\cite{Peres2} We note that
the effective Hamiltonian given above is valid only for wave vectors $k\ll\Lambda$, where $%
\Lambda\simeq 8.25$ eV is a high-energy cutoff stemming from the
discreteness of the lattice.\cite{Peres2}

The quantity of interest for many physical properties is the dynamical
polarization, since it determines e.g. the effective electron-electron
interaction, the Friedel oscillations and the plasmon and phonon spectra. In
terms of the bosonic Matsubara frequencies $\omega_n=2\pi n/\beta$, it is
defined as\cite{Mahan}
\begin{align}
P(\mathbf{q},i\omega_n)&=-\frac{1}{A}\int_0^\beta d\tau
e^{i\omega_n\tau}\langle T\rho(\mathbf{q},\tau)\rho(-\mathbf{q},0)\rangle \,,
\end{align}
where $A$ denotes the area. The average is taken over the canonical ensemble
and the density operator is given by the sum of the density operators of the
two sub-lattices $\rho=\rho_a+\rho_b$. This amounts to working in the
long-wavelength limit. To first order in the electron-electron interaction,
we obtain
\begin{align}
P^{(1)}(\mathbf{q},i\omega_n)&=\frac{g_S g_V}{4\pi^2}\int
d^2k\sum_{s,s^{\prime}=\pm}f^{ss^{\prime}}(\k,\mathbf{q})\,
\frac{n_F(E^{s}(k))-n_F(E^{s^{\prime}}(|\mathbf{k} + \mathbf{q}|))}{%
E^{s}(k)-E^{s^{\prime}}(|\mathbf{k}+\mathbf{q}|)+i\hbar \omega_n}\,,
\label{eq:P1}
\end{align}
with $E^{\pm}(k)=\pm \hbar v_F k-\mu$ the eigenenergies, $n_F(E)=(e^{\beta
E}+1)^{-1}$ the Fermi function, and $g_S=g_V=2$ the spin and valley
degeneracy. A characteristic difference between the polarization of graphene
and that of a 2DEG is the appearance of the prefactors $f^{ss^{\prime}}(\k,%
\mathbf{q})$ coming from the band-overlap of the wave functions\cite%
{Shung,Ando06}
\begin{align}
f^{s s^{\prime}}(\k,\mathbf{q})&=\frac{1}{2}\left(1+s s^{\prime}\frac{%
k+q\cos\varphi}{|\k+\mathbf{q}|}\right),
\end{align}
where $\varphi$ denotes the angle between $\k$ and $\mathbf{q}$.

At zero temperature, the Fermi functions yield simple step functions. We
define the following retarded function by replacing $i\omega _{n}\rightarrow
\omega +i\delta $%
\begin{align}
\chi _{D}^{\pm}(\mathbf{q},\omega )& =\frac{g}{4\pi ^{2}\hbar
}\int_{k\leq D}d^{2}k\sum_{\alpha =\pm }  \, \frac{\alpha f^{\pm
}(\mathbf{k},\mathbf{q})}{\omega +\alpha v_{F}(k\mp
|\mathbf{k}+\mathbf{q}|)+i\delta }\,\,, \label{DefSusc}
\end{align}%
where $g\equiv g_{S}g_{V}$. The $+(-)$ sign corresponds to intra(inter)-band
transitions and $D$ is a general upper limit.

For $\mu =0$, the retarded polarization thus reads
\begin{equation}
P_{0}^{(1)}(\mathbf{q},\omega )=-\chi _{\Lambda
}^{-}(\mathbf{q},\omega )\label{eq:P0}.
\end{equation}%
For $\mu >0$, i.e., for nonzero electron doping, the retarded polarization
has an additional term
\begin{equation}
\Delta P^{(1)}(\mathbf{q},\omega )=\chi _{\mu
}^{+}(\mathbf{q},\omega )+\chi _{\mu }^{-}(\mathbf{q},\omega
)\label{eq:Pmu} .
\end{equation}

In the Appendix we give additional details of the calculation of Eq.~(\ref%
{DefSusc}). Here we summarize these expressions in terms of two complex
functions, $F(q,\omega )$ and $G(x)$, defined as
\begin{align}
F(q,\omega )& =\frac{g}{16\pi }\frac{\hbar
v_{F}^{2}q^{2}}{\sqrt{\omega ^{2}-v_{F}^{2}q^{2}}}\,,\quad
G(x)=x\sqrt{x^{2}-1}-\ln \left( x+\sqrt{x^{2}-1}\right)
\label{eq:complexG} \,.
\end{align}%
From now on it is always assumed that $\omega >0$, noting that the
polarization for $\omega <0$ is obtained via $P^{(1)}(\mathbf{q},-\omega )=%
\left[ P^{(1)}(\mathbf{q},\omega )\right] ^{\ast }$. Equations
(\ref{eq:P0}) and (\ref{eq:Pmu}) are then rewritten in the following
compact form
\begin{equation}
P^{(1)}(\mathbf{q},\omega )=P_{0}^{(1)}(\mathbf{q},\omega )+\Delta P^{(1)}(%
\mathbf{q},\omega )\label{eq:Pfinal}\,,
\end{equation}%
with
\begin{equation}
P_{0}^{(1)}(q,\omega )=-i\pi \frac{F(q,\omega )}{\hbar ^{2}v_{F}^{2}}\,,
\end{equation}%
and
\begin{align}
\Delta P^{(1)} (q,\omega )=&-\frac{g\mu }{2\pi \hbar
^{2}v_{F}^{2}}+\frac{ F(q,\omega )}{\hbar ^{2}v_{F}^{2}}\left\{
G\left( \frac{\hbar \omega +2\mu }{\hbar v_{F}q}\right) -\Theta
\left( \frac{2\mu -\hbar \omega }{\hbar v_{F}q}-1\right) \left[
G\left( \frac{2\mu -\hbar \omega }{\hbar v_{F}q}\right) -i\pi
\right] \right.
\notag \\
& \left. -\Theta \left( \frac{\hbar \omega -2\mu }{\hbar
v_{F}q}+1\right) G\left( \frac{\hbar \omega -2\mu }{\hbar
v_{F}q}\right) \right\} \label{eq:DP} .
\end{align}

Equations~(\ref{eq:Pfinal})-(\ref{eq:DP}) are the main result of this work.
Details of the calculation are given in the Appendix, where we also give
expressions for the real and imaginary part of the polarization in terms of
real functions. 

Two limits of the polarization are of particular importance: (i) {The long
wavelength limit $q\rightarrow 0$ with $\omega >v_{F}q$ fixed, which is
relevant for optical spectroscopy and for the plasma dispersion. (ii) The
static case $\omega =0$ with q arbitrary, which is relevant for the
screening of charged or magnetic impurities.}

For the first case we obtain
\begin{align}
P^{(1)}(q\to0,\omega)=\frac{g q^2}{8\pi \hbar \omega}\Big[&\frac{2\mu}{\hbar
\omega}+\frac{1}{2}\ln\left|\frac{2\mu-\hbar \omega}{2\mu+\hbar \omega}%
\right|  -i\frac{\pi}{2}\Theta(\hbar \omega-2\mu)\Big].
\label{LongWave}
\end{align}

And for the second case we recover previous results~\cite{Ando06}
\begin{align}
P^{(1)}(q,0) &=-\frac{g k_F}{2\pi \hbar v_{F}}   +\Theta (q-2k_F)\frac{gq}{8\pi \hbar v_{F}}G_{<}\left( \frac{2k_F }{q}\right) ,
\end{align}%
where $G_{<}(x)\equiv -iG(x)$ is the real function (for $|x|<1$)
given in Eq.~(\ref{eq:realG}). Note that for $2\mu \leq \hbar
v_{F}q\leq \Lambda $ the absolute value of
the polarization is linear in $q$, and acquires rapidly the behavior of $%
P_{0}^{(1)}$. By contrast, the polarization of the ordinary 2DEG decreases
with increasing wave vector for $q>2k_{F}$. Furthermore, and also in
contrast to the 2DEG, where the first derivative of the static polarization
is discontinuous at $q=2k_{F}$, doped graphene has a continuous first
derivative at $\hbar v_{F}q=2\mu $ and a discontinuous second derivative.

Table~\ref{tab:Im} summarizes the $\omega $ dependence of $\left\vert \text{%
Im}P^{(1)}\right\vert $ for $\w\rightarrow 0$ for a 2DEG and for
doped graphene. While the dependence is the same for $q\neq 2 k_F$, a subtle difference appears at $q=2 k_F$. This difference is intimately related to the nonanalytic behavior
(as a function of $q$) of the static polarization at $q=2 k_F$. It leads to e.g. a different power law decay of the
electron screening, as will be shown in subsection~\ref{Friedel}.
Table~\ref{tab:Im} also indicates that, depending on the order in which the limits $%
q,\mu \rightarrow 0$ are taken, the low-frequency behavior may be
that of an insulator, a metal, or a hybrid between the two.

\begin{figure}[t]
\begin{center}
\includegraphics[angle=0,width=0.6\linewidth]{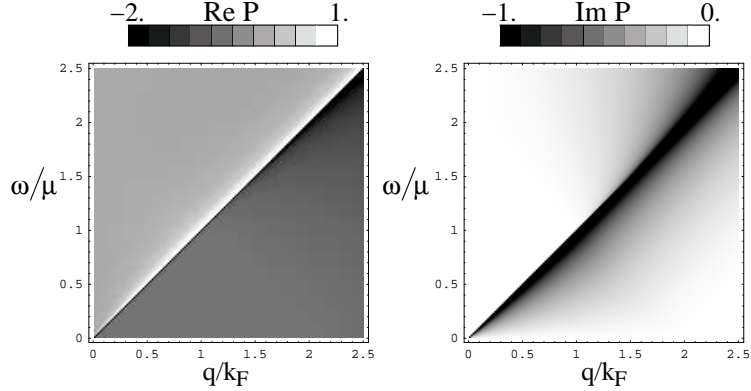}
\end{center}
\caption{Density plot of $\text{Re} P^{(1)}(q,\protect\omega)$
(left)
and $\text{Im} P^{(1)}(q,\protect\omega)$ (right) in units of $%
\protect\mu/\hbar^2 v_F^2$. We set $\hbar=v_F=1$. }
\label{fig:P}
\end{figure}

The polarization is a continuous function of $q$ and $\omega $ except for
the square-root divergence which $F(q,\omega )$ shows at $\omega =v_{F}q$.
Figure~\ref{fig:P} shows real and imaginary parts of the polarization given
by Eq.~(\ref{eq:Pfinal}). We note that Im$P^{(1)}(q,\omega )=0$ for $\hbar
v_{F}q<\hbar \omega <2\mu -\hbar v_{F}q$ or $0<\hbar \omega <\hbar
v_{F}q-2\mu $, and negative otherwise, while Re$P^{(1)}(q,\omega )<0$ for $%
\omega <v_{F}q$.

\begin{figure}[t]
\begin{center}
\includegraphics[angle=0,width=0.6\linewidth]{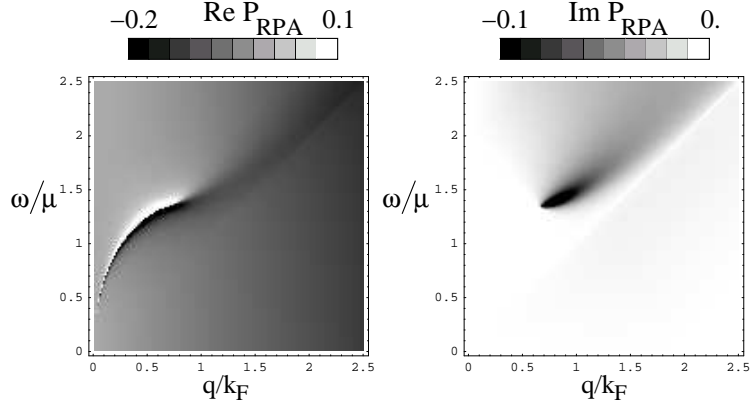}
\end{center}
\caption{Same as Fig.~\protect\ref{fig:P}, for the renormalized polarization
$P_{\mathrm{RPA}}(q,\protect\omega)$. }
\label{fig:PRen}
\end{figure}

The divergence at $\omega =v_{F}q$ vanishes in the self-consistent RPA
result of the polarization given by~\cite{Fetter}
\begin{equation}
P_{\mathrm{RPA}}(\mathbf{q},\omega )=\frac{P^{(1)}(\mathbf{q},\omega )}{%
1-v_{q}P^{(1)}(\mathbf{q},\omega )}~.
\end{equation}%
Here $v_{q}=e^{2}/2\kappa _{0}q$ denotes the in-plane Coulomb potential in
vacuum. We note that $P_{\mathrm{RPA}}(q,v_{F}q)=-v_{q}$ so that the
self-consistent polarization has a real and finite value at $\omega =v_{F}q$%
. Figure~\ref{fig:PRen} shows real and imaginary parts of the
self-consistent polarization. While the singularity at $\omega =v_{F}q$ is
absent, a new singularity appears in $\text{Re}P_{\mathrm{RPA}}(q,\omega )$
at $\omega \propto \sqrt{q}$, which reflects the existence of plasmons, as
will be discussed in subsection~\ref{plasmon}.

\begin{table}[tbp]
\begin{tabular}{|l|c|c|c|}
\hline & $y<1$ & $y=1$ & $y>1$ \\ \hline 2DEG & $\omega$ &
\textbf{$\omega^{1/2}$} & 0 \\ \hline graphene & $\omega$ &
\textbf{$\omega^{3/2}$} & 0 \\ \hline
\end{tabular}%
\caption{Low-frequency dependence of $|\mathrm{Im}P^{(1)}|$ in the limit $%
\w\to0$ for a 2DEG and for doped graphene. Here $y\equiv q/2 k_F$.} \label{tab:Im}
\end{table}

\section{Static screening}

\label{Friedel} 
An external charge density $n^{\mathrm{ext}}(\mathbf{r})=Ze\delta (\mathbf{r}%
)$ is screened by free electrons due to the Coulomb interaction. This
results in the induced charge density $\delta n(r)$
\begin{equation}
\delta n(r)=\frac{Ze}{4\pi ^{2}}\int d^{2}q\left[ \frac{1}{\epsilon (q,0)}-1%
\right] e^{i\mathbf{q\cdot }\mathbf{r}}\,.
\end{equation}

Here $\epsilon (q,0)\equiv \lim_{\omega \rightarrow 0}\epsilon
(q,\omega )$. Within the RPA approximation \cite{Shung}
\begin{equation}
\epsilon (q,\omega )={\epsilon _{0}}-v_{q}P^{(1)}(q,\omega
)\label{eq:eel}~.
\end{equation}%
The effective dielectric constant $\epsilon _{0}$ includes high energy
screening processes. We take $\epsilon _{0}\simeq 2.4$.\cite{Ando06}

There are two contributions to the induced charge density. A non-oscillating
part comes from the long-wavelength behavior of the polarization. This
contribution is obtained within the Thomas-Fermi (TF) approximation and is
given by
\begin{equation}
\delta n_{\mathrm{TF}}(r)\simeq
           -\frac{Ze}{2\pi \alpha g k_F r^{3}}
\label{eq:TF}%
\,,
\end{equation}%
where $\alpha \equiv e^{2}/4\pi \kappa _{0}\hbar v_{F}\simeq 2.5$. We note
here that in a 2DEG the TF contribution also decays as $r^{-3}$.\cite%
{Stern67}

\begin{figure}[t]
\begin{center}
\includegraphics[angle=-90,width=0.6\linewidth]{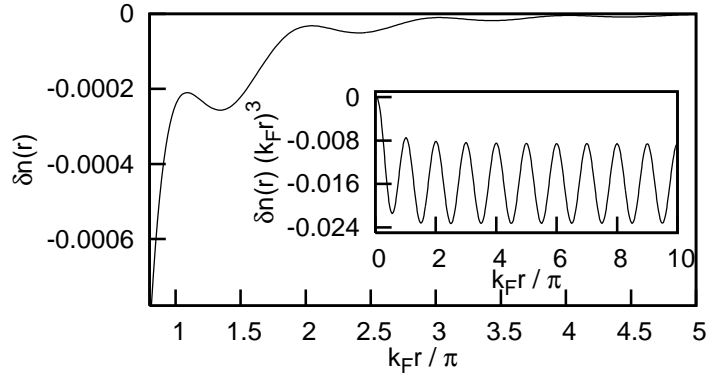}
\end{center}
\caption{Induced charge density $\protect\delta n(r)$ (in units of $\protect%
k_F^2 Z e$) as a function of the dimensionless variable $k_F r/\pi$. The inset shows the $r^{-3}$ decay at large distances. The high
frequency screening is set to $\protect\epsilon_0=2.4$. In the used
representation the graphs are invariant under a change of the chemical
potential.}
\label{fig:Friedel}
\end{figure}

The second contribution to the long distance behavior is oscillatory and
comes from the non-analyticity of the polarization at $\hbar v_{F}q=2\mu $.%
\cite{Lighthill,Stern67} However, and quite importantly, in graphene
the non-analyticity results from a discontinuity occurring only in
second derivative, the first derivative being continuous. This leads
to an oscillatory decay
\begin{equation}
\delta n_{\mathrm{osc}}(r)\propto \frac{Ze \cos (2 k_F r)}{k_F (\epsilon _{0}+2\alpha
             )^{2}r^{3}}\,\label{eq:osc},
\end{equation}%
which contrasts with the behavior of a 2DEG, where Friedel
oscillations scale like $\delta n(r)\propto \cos (2k_{F}r)\,r^{-2}$.
This difference has been previously noted in Refs.
\onlinecite{Cheianov,Katsnelson1}, where however TF screening was
not considered. We emphasize here that, because the TF contribution
is of the same order of magnitude as the oscillatory part, it is
essential to consider both of them. Furthermore we note that, while
the TF contribution to screening
given in Eq.~(%
\ref{eq:TF}) is independent of the dielectric constant $\epsilon
_{0}$, the amplitude of the oscillatory part given in
Eq.~(\ref{eq:osc}) decreases with increasing $\epsilon _{0}$. In
fact, our numerical calculations show that for large distances the
induced density does not change sign even in the hypothetical case
of $\epsilon _{0}=0$ where the ratio between oscillatory and TF
contribution would be maximal. This remarkable general property can
be clearly appreciated in Fig.~\ref{fig:Friedel} for the particular
case of $\epsilon _{0}=2.4$: The induced density $\delta n(r)$ does
not change sign, but oscillates around a finite offset.

\begin{figure}[t]
\begin{center}
\includegraphics[angle=-90,width=0.6\linewidth]{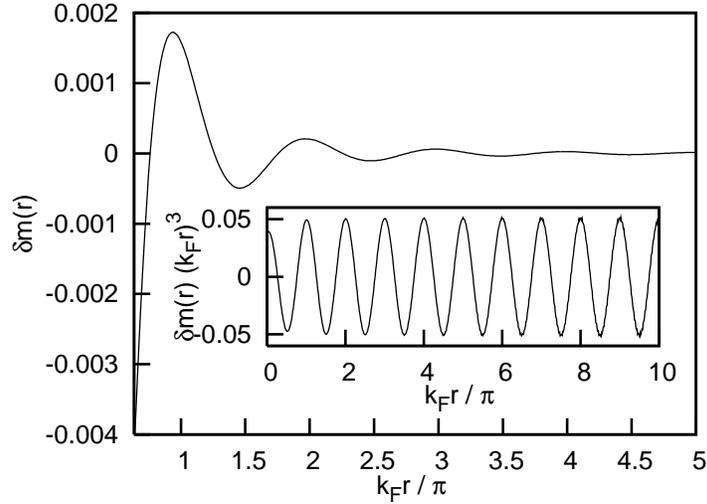}
\end{center}
\caption{Same as Fig.~\protect\ref{fig:Friedel}, for the induced spin
density $\protect\delta m(r)$. }
\label{fig:RKKY}
\end{figure}

The polarization also determines the Ruderman-Kittel-Kasuya-Yosida (RKKY) interaction energy between two
magnetic impurities as well as the induced spin density due to a magnetic
impurity, both quantities being proportional to the Fourier transform of $%
P^{(1)}(q,0)$.\cite{RKKY} For a magnetic impurity at $\mathbf{r}$
the induced spin density $\delta m(\mathbf{r})$ is
\begin{equation}
\delta m(\mathbf{r})\propto P^{(1)}(\mathbf{r})=\frac{1}{4\pi
^{2}}\int d^{2}q\,P^{(1)}(q,0)e^{i\mathbf{q}\cdot\mathbf{r}}\,.
\end{equation}%
Note that the magnetic response is proportional to the bare polarization, since the spin-spin interaction is mediated via the short ranged exchange interaction (in contrast to the long ranged Coulomb interaction in the case of charge response). That is why the TF
contribution is missing and the induced spin density oscillates around zero. Figure~\ref{fig:RKKY} shows the numerically
calculated $P^{(1)}(r)$ in units of $k_F^3/\hbar v_F$ and illustrates this behavior. Specifically in the
long wavelength limit we obtain
\begin{equation}
\delta m(r)\propto \frac{\cos (2 k_F r)}{r^{3}}\,,
\end{equation}%
which is clearly seen in the inset of Fig~\ref{fig:RKKY}. Like for the
induced charge density, we find that the induced spin polarization $\delta
m(r)$ decreases like $r^{-3}$ for large distances. Again, this contrasts
with the $r^{-2}$ behavior found in a 2DEG.\cite{RKKY} For the particular
case of undoped graphene we recover the monotonous $r^{-3}$ decay obtained
in Ref.~\onlinecite{Geli}.

Finally we note, that in contrast to the behavior of charge screening, where
the induced charge scales like $1/\mu $ at large distances, the envelop
function of the spin density is independent of the chemical potential at
large distances.

\section{Plasmons}

\label{plasmon} 
\begin{figure}[t]
\begin{center}
\includegraphics[angle=0,width=0.6\linewidth]{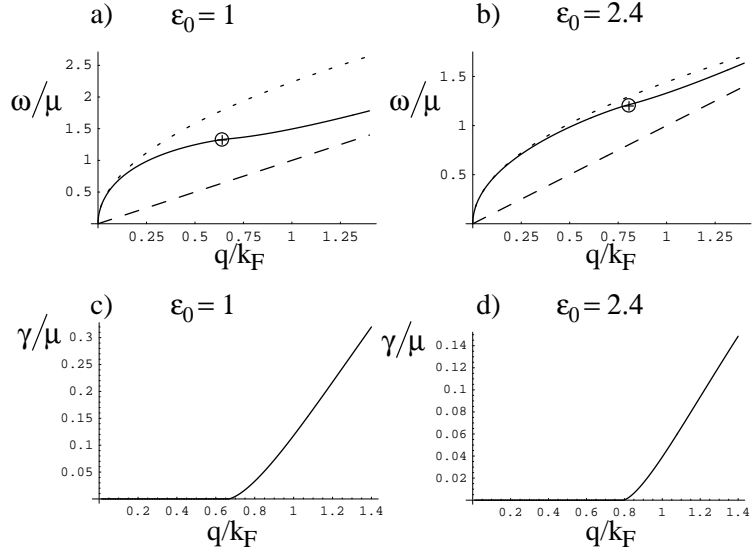}
\end{center}
\caption{Upper row: Solid lines show the dispersion relation for plasmons
defined by Re $\protect\epsilon(q,\protect\omega)=0$. Dotted lines show the
low-q expansion of Eq.~(\protect\ref{eq:plasma}), while the dashed lines
represents $\protect\omega=v_F q$. The crosses indicate where the plasmons
acquire a finite lifetime. Lower row: Decay rate $\protect\gamma$ of
plasmons [see Eq.~(\protect\ref{eq:gamma})] in units of chemical potential.
We set $\hbar=1$.}
\label{fig:Plasmon}
\end{figure}

The plasmon dispersion is determined by solving for $\epsilon (q,\omega
_{p}-i\gamma )=0$, where $\gamma $ is the decay rate of the plasmons. For
weak damping, the plasmon dispersion $\omega _{p}(q)$ and the decay rate $%
\gamma $ are determined by\cite{Fetter}
\begin{align}
\epsilon _{0}& =v_{q}\text{Re}P^{(1)}(q,\omega _{p})\,, \quad\gamma
 =\frac{\text{Im}P^{(1)}(q,\omega _{p})}{\frac{\partial }{\partial
\omega }\text{Re}P^{(1)}(q,\omega )\left. {}\right\vert _{\omega
_{p}}}. \label{eq:gamma}
\end{align}%
Solutions to the first equation only exist for Re$P^{(1)}>0$, which is the
case only for finite doping and $\omega > v_F q$. Furthermore, a stable solution
requires Im$P^{(1)}=0$, as is the case for region 1B of Fig.~\ref%
{fig:Regions}. Using the low-$q$ expansion of the polarization given in Eq.~(%
\ref{LongWave}), and neglecting the logarithmic correction, we obtain
\begin{equation}
\hbar \omega _{p}(q)=\left( \frac{g\alpha \mu \hbar v_{F}q}{2\epsilon _{0}}%
\right)^{1/2}\label{eq:plasma},
\end{equation}%
with $\alpha $ defined below Eq.~(\ref{eq:TF}). The $\sqrt{q}$ behavior of
the plasmon dispersion also appears in the 2DEG.


Outside region 1B, the plasmon is damped, i.e. it has a nonzero decay rate $%
\gamma $. This can be clearly seen in Figs. \ref{fig:PRen} and ~\ref%
{fig:Plasmon}. In the upper row of Fig.~\ref{fig:Plasmon}, we plot the exact
plasmon dispersion for two different values of $\epsilon _{0}$ and indicate
the point at which the collective excitation becomes damped. The lower row
shows the decay rate as obtained from Eq.~(\ref{eq:gamma}).

Finally, it is interesting to note that the combination of the
linear dispersion relation for quasiparticles [electrons above (or
holes below) the Fermi energy] and the plasmon dispersion makes it
impossible for a quasiparticle of energy $\hbar \omega$ to decay
into a plasmon with $q \leq \omega/v_F$. Hence, plasmons with
infinite lifetime do not contribute to the lifetime of
quasiparticles.

\section{Acoustic phonons and sound velocity}

We now calculate the dispersion and the decay rate of acoustical
phonons in graphene. We treat the electrons in the $\pi $-band of
graphene as quasi-free electrons, while all the other electrons are
assumed to be tightly bound to the carbon nuclei, thus forming
effective ions with a positive elementary charge. In the absence of
screening by the conduction electrons, the ions oscillate at their
plasma frequency due to the long range nature of the Coulomb
interaction. The collective modes of the combined electron-ion
plasma can be obtained from the zeros of the total dielectric
function $\epsilon _{%
\mathrm{tot}}(q,\omega )$, which is obtained by summing the
contributions from ions and electrons:~\cite{Ashcroft}
\begin{align}
\epsilon _{\mathrm{tot}}(q,\omega )& =\epsilon _{\mathrm{el}}(q,\omega
)+\epsilon _{\mathrm{ion}}(q,\omega )-1  =\epsilon _{0}-v_{q}\left[ P_{\mathrm{el}}(q,\omega )+P_{\mathrm{ion}%
}(q,\omega )\right]  \label{eq:etot}
\end{align}%
Here $P_{\mathrm{el}},\epsilon _{\mathrm{el}}$ are the polarization and the
dynamical dielectric function of the electrons as given in Eqs.~(\ref%
{eq:Pfinal}) and (\ref{eq:eel}), while $P_{\mathrm{ion}},\epsilon _{\mathrm{%
ion}}$ are the corresponding quantities for the ions. In the calculation of $%
P_{\mathrm{ion}}(q,\omega )$ we assume that the ions have a
quadratic energy dispersion $E=\hbar ^{2}k^{2}/2M$ where $M$ denotes
the ion mass. The ionic charge density is two positive charges per
unit cell, so that the Fermi wave vector of the ions is
$k'_{F}\simeq \left( 8\pi /A_{c}\right)^{1/2}$, where
$A_{c}=3\sqrt{3}a^{2}/2$ denotes the area of the hexagonal unit cell
in real space.\cite{Peres2} This value of $k'_{F}$ is exact for
field effect doping and approximate for chemical doping.

In the calculation of $P_{\mathrm{ion}}$ we assume that, for all
relevant frequencies, we can take $\omega \gg \hbar k'_{F}q/M$. In
the case of acoustic phonons this assumption is equivalent to
$v_{s}\gg \hbar k'_{F}/M\simeq 1.3\times 10^{-4}v_{F}$, where
$v_{s}$ denotes the sound velocity. This relation is fulfilled for
all meaningful dopings as shown below. In this regime the ion
polarization is real and given by
\begin{equation}
P_{\mathrm{ion}}(q,\omega )=\frac{{k'_{F}}^{2}q^{2}}{4\pi M\omega ^{2}}=\frac{%
2E_{0}q^{2}}{\hbar ^{2}\omega ^{2}}\,,
\end{equation}%
where we defined $E_{0}=\hbar ^{2}/MA_{c}\simeq 7\times 10^{-5}$ eV. We note
that $E_{0}$ is of the order of the ion confinement energy.

We may estimate the dispersion and the decay rate of the acoustical phonons
by inserting Eq.~(\ref{eq:etot}) into Eq.~(\ref{eq:gamma}). For finite
doping $\mu >0$, the acoustic phonons at long wavelengths lie in the region
1A of Fig.~\ref{fig:Regions} (defined by $\omega <v_{F}q<2\mu /\hbar -\omega
$), where $\text{Re}P^{(1)}(q,\omega )=-g\mu /2\pi \hbar ^{2}v_{F}^{2}$. In
this regime the phonon dispersion is easily obtained:
\begin{equation}
\omega _{\mathrm{ph}}=\left( \frac{4\pi \a E_{0}}{\epsilon _{0}\hbar v_{F}q+g%
\a\mu }\right) ^{1/2}v_{F}q\label{eq:wph}\,.
\end{equation}%
To be consistent with the precondition $\omega <v_{F}q$ the expression in
the square root has to be smaller than one. This sets a lower limit to the
values of the chemical potential for which Eq.~(\ref{eq:wph}) is valid. The
sound velocity $v_{s}$ and the decay rate $\gamma $ may be derived from Eqs.
(\ref{eq:gamma}) and~(\ref{eq:wph}) in the limit of low $q$. We obtain%
\begin{equation}
v_{s}=\sqrt{\xi }\;v_{F}\,,\quad \gamma =\frac{\xi
v_{F}q}{2\sqrt{1-\xi }}\label{eq:vs}\,,
\end{equation}%
with $\xi =4\pi E_{0}/g\mu $.

Some additional remarks on the validity of Eq.~(\ref{eq:wph}) go in
place. We have already said that, for $q\rightarrow 0$, it only
applies provided $\xi <1$. We also wish to note that the acoustical
phonons are only well-defined if
their frequency is much larger than their decay rate, i.e. if $\omega _{%
\mathrm{ph}}/\gamma \gg 1$. Since $\omega _{\mathrm{ph}}/\gamma =1$ for $\xi
=4/5$, we conclude that the notion of acoustic phonons is justified for $\xi
\ll 1$, which corresponds to $\mu \gg 4\pi E_{0}/g\simeq 2.2\times 10^{-4}$
eV. A detailed discussion of acoustical phonons for $\xi >1$ is left for
future work.

We have assumed initially that $v_{s}\gg \hbar k'_{F}/M\simeq
1.3\times 10^{-4}v_{F}$. According to Eq.~(\ref{eq:vs}) this results
in $\xi \gg 1.7\,\times 10^{-8}$, or equivalently, $\mu \ll
10^{8}E_{0}$, which is always fulfilled.

We finish this section by estimating the sound velocity of a typical
graphene sample. Assuming a concentration of "conduction-band" electrons of $%
n_{\mathrm{el}}=10^{10}-10^{12}$ $\mathrm{cm^{-2}}$ we get $\mu
\simeq 10^{-2}-10^{-1}$ eV. We note that this corresponds to $\xi
\simeq 2-20\times 10^{-3}$, so that Eqs.~(\ref{eq:wph})
and~(\ref{eq:vs}) are applicable, resulting in $v_{s}\simeq
0.05-0.14\,v_{F}\simeq 4-12\times 10^{4}$ m/s and $\gamma
\simeq10^{-3}-10^{-2}\,v_{F}q$. These results suggest a significant
enhancement of the sound velocity, as compared to normal metals,
where $v_s \simeq \sqrt{m/2M}v_F\alt 10^{-2}v_F$. The low
polarizability of the conduction electrons leads to a poor screening
of the oscillations of the charged ions. On the other hand, in a
semiconductor with a gap larger than the typical acoustic phonon
frequencies, the electrons follow adiabatically the ions. In that
case, the lattice vibrations can be described as oscillations of
neutral particles.

\section{Conclusions}


In this article we have derived a compact and closed expression for the
dynamical polarization of graphene within the RPA approximation. The
obtained result is valid for arbitrary wave vector, frequency, and doping.
As particular cases, we have derived the long-wavelength limit $q\to0$ and
the static limit $\omega\to 0$. We have employed the RPA polarization to
calculate several physical quantities of interest in doped graphene. First
we have studied the static Friedel oscillations of the induced charge(spin)
density in the presence of a charged(magnetic) impurity. We have found that,
although the charge density does show oscillations around an average value,
it does it without changing sign. The reason for this remarkable behavior is
that Friedel oscillations superpose on the dominant Thomas-Fermi induced
density, with both contributions decaying at long distances $r$ with the
same power law $r^{-3}$.

The dynamical polarization has been used to calculate the dispersion
relation and the decay rate of plasmons and acoustic phonons. Like in the
2DEG case, the plasmon frequency shows a $\sqrt{q}$-behavior in the long
wavelength regime. We have determined the region in the $(q,\omega)$ plane
where the plasmon is stable, as well as the decay rate in the regime where
it is not.

The dispersion of acoustical phonons has been shown to be strongly
dependent on the chemical potential. In particular, we have found
that the sound velocity approaches the Fermi velocity at low doping.
However, the same limit shows an increase in the decay rate of
acoustic phonons due to electron-hole pair excitation.

Although we have focused on applications of the RPA calculation to
the case of doped graphene, some aspects of the low-frequency,
long-wavelength dynamics of pure graphene appear to be intriguing
and worth studying further.

{\it Note added.} When this work was about to be submitted, we
became aware of a related paper.\cite{hwang06} Overlapping results
in the two papers are in agreement.

\section*{Acknowledgements}

We appreciate helpful discussions with A. H. Castro Neto. This work
has been supported by the EU Marie Curie RTN Programme No.
MRTN-CT-2003-504574, the EU Contract 12881 (NEST), and by MEC
(Spain) through Grants No. MAT2002-0495-C02-01,
FIS2004-05120, FIS2005-05478-C02-01, the Juan de la
Cierva Programme and the Comunidad de Madrid program CITECNOMIK, ref. CM2006-S-0505-ESP-0337.
\appendix

\section{Calculation of the polarization}

\begin{figure}[t]
\begin{center}
\includegraphics[angle=0,width=0.5\linewidth]{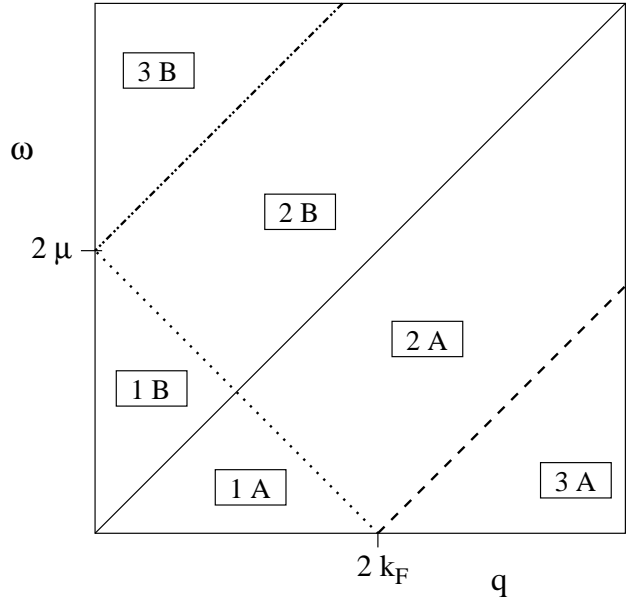}
\end{center}
\caption{Display of the different regions characterizing the susceptibility
behavior. Regions are limited by straight lines $\omega=v_F q$ (solid), $%
\protect\omega=v_F q-2\protect\mu$ (dashed) and $\protect\omega=2\protect\mu-v_F q$
(dotted), where we set $\hbar=1$. }
\label{fig:Regions}
\end{figure}
In the following we present some major steps of the calculation of the
polarization. We restrict the discussion to $\omega>0$ since $%
P^{(1)}(q,-\omega)=\left[P^{(1)}(q,\omega)\right]^*$. In all the Appendix we
set $v_F=\hbar=1$, so that in
             the Appendix $\mu=k_F$.

\subsection{Imaginary part}

The imaginary part of the functions $\chi_{D}^\pm(q,\omega)$ defined in Eq.~(%
\ref{DefSusc}) has the following form
\begin{align}
\text{Im}\chi_{D}^\beta(q,\omega)= -\frac{g}{4\pi}\int_{0}^{D} dk
\sum_{\alpha=\pm} \alpha\, I^{\a\b}(k,q,\omega)\,,  \notag \\
I^{\a\b}=k\,\int_{0}^{2\pi}\,d\varphi\, f^\beta(\k,\mathbf{q})\delta\left[%
\omega+\alpha(k-\beta|\k+\mathbf{q}|)\right] \,,  \notag
\end{align}
The $\varphi$-integration yields
\begin{align}
I^{\a\b}= &\left[\frac{(2\a k +\omega)^2-q^2}{q^2-\omega^2}\right]^\frac{1}{2}%
\left\{\Theta(\beta)\Theta(q-\omega)\Theta\left(k-\frac{q-\alpha \omega}{2}%
\right)\right.  \notag \\
&\left.+\Theta(-\beta)\Theta(\omega-q)\Theta(-\alpha)\left[\Theta\left(\frac{%
\omega+q}{2}-k\right)-\Theta\left(\frac{\omega-q}{2}-k\right)\right]\right\}
,  \notag
\end{align}
which is always real. The final $k$-integration can now simply be
performed. We obtain for $\mu =0$
\begin{align}
\text{Im}P_{0}^{(1)}(q,\omega )& =\frac{g}{4\pi }\int_{0}^{\Lambda
}dk\sum_{\alpha }\alpha \,I^{\a\,-}(k,q,\omega )
=-\frac{gq^{2}}{16\sqrt{\omega ^{2}-q^{2}}}\Theta (\omega -q).
\label{eq:Im0}
\end{align}%
In order to present the result for $\mu >0$, we introduce the real functions $%
f(q,\omega ),G_{>}(x),G_{<}(x)$
\begin{align}
f(q,\omega )& =\frac{g}{16\pi }\frac{q^{2}}{\sqrt{|\omega ^{2}-q^{2}|}}\,,
\notag \\
G_{>}(x)& =x\sqrt{x^{2}-1}-\cosh ^{-1}(x)\,, \quad x>1 \, , \notag \\
G_{<}(x)& =x\sqrt{1-x^{2}}-\cos ^{-1}(x)\, , \quad |x|<1\, .
\label{eq:realG}
\end{align}%
For the additional term at finite doping given by Eq.~(\ref{eq:Pmu}), we
obtain in the language of Fig.~\ref{fig:Regions}
\begin{align}
\text{Im}\Delta P^{(1)}(q,\omega )&=-\frac{g}{4\pi }\int_{0}^{\mu
}dk\sum_{\alpha ,\beta }\alpha \,I^{\a\beta }(k,q,\omega )
=f(q,\omega )\times\left\{
\begin{array}{ll}
G_{>}(\frac{2\mu -\omega }{q})-G_{>}(\frac{2\mu +\omega }{q}) & ,\text{ 1 A}
\\[1.5ex]
\pi & ,\text{ 1 B} \\
-G_{>}(\frac{2\mu +\omega }{q}) & ,\text{ 2 A} \\
-G_{<}(\frac{\omega -2\mu }{q}) & ,\text{ 2 B} \\
0 & ,\text{ 3 A} \\
0 & ,\text{ 3 B}
\end{array}%
\right.  \notag
\end{align}

\subsection{Real part}

The Kramers-Kronig relation valid for the retarded function $\text{Re}%
P^{(1)}_0(q,\omega)$ reads
\begin{align}
\text{Re}P^{(1)}_0(q,\omega)&=\frac{1}{\pi}\displaystyle\int\limits_{-%
\infty}^{\infty} d\omega^{\prime}\,\frac{\text{Im}P^{(1)}_0(q,\omega^{%
\prime})}{\omega^{\prime}-\omega} =-\frac{ g q^2}{16
\sqrt{q^2-\omega^2}}\Theta(q-\omega) .  \label{eq:Re0}
\end{align}

For finite doping we rewrite Eq.~(\ref{eq:Pmu}) as
\begin{align}
\text{Re}\Delta P^{(1)}(q,\omega )& =\frac{g}{4\pi ^{2}}\int_{0}^{\mu
}dk \, k\int_{0}^{2\pi }d\varphi  \, \sum_{\alpha =\pm }\frac{2k+\alpha \omega +q\cos \varphi }{%
(k+\alpha \omega )^{2}-|\mathbf{k}+\mathbf{q}|^{2}}\quad .  \notag
\end{align}

This integral is calculated directly such that the Kramers-Kronig relation
is not needed, here. The $\varphi $-integration yields
\begin{equation}
\text{Re}\Delta P^{(1)}=-\frac{g\mu }{2\pi }+\frac{g}{8\pi ^{2}}%
\sum_{\alpha =\pm }\int_{0}^{\mu }dkJ^{a}(k,q,\omega ),
\end{equation}

where $J^\a(k,q,\omega)$ is given by
\begin{align}
J^\a=&2\pi \left[\frac{(2\a k +\omega)^2-q^2}{\omega^2-q^2}\right]^\frac{1}{2%
}\left\{\Theta(q-\omega)\Theta(\frac{q-\alpha \omega}{2}-k)\right.  \notag \\
&\left.+\Theta(\omega-q)\left[\Theta(\alpha)+\Theta(-\alpha)\left(\Theta(%
\frac{\omega-q}{2}-k)-\Theta(k-\frac{\omega+q}{2})\right)\right]\right\} .
\notag
\end{align}

We thus get in the language of Fig.~\ref{fig:Regions}
\begin{align}
\text{Re}& \Delta P^{(1)}(q,\omega )=-\frac{g\mu }{2\pi }
+f(q,\omega )\times\left\{
\begin{array}{ll}
\pi & ,\text{ 1 A} \\[1.3ex]
-G_{>}(\frac{2\mu -\omega }{q})+G_{>}(\frac{2\mu +\omega }{q}) & ,\text{ 1 B}
\\[1.3ex]
-G_{<}(\frac{\omega -2\mu }{q}) & ,\text{ 2 A} \\[1.3ex]
G_{>}(\frac{2\mu +\omega }{q}) & ,\text{ 2 B} \\[1.3ex]
-G_{<}(\frac{\omega -2\mu }{q})+G_{<}(\frac{2\mu +\omega }{q}) & ,\text{ 3 A}
\\[1.3ex]
G_{>}(\frac{2\mu +\omega }{q})-G_{>}(\frac{\omega -2\mu }{q}) & ,\text{ 3 B}
\end{array}%
\right.  \notag
\end{align}

\subsection{Analytic representation}

Concerning the analytic representation of the results given in Eq.~(\ref%
{eq:Pfinal}), we note the following properties (see e.g. Ref~%
\onlinecite{math})
\begin{align}
\Theta(x-1) \cosh^{-1}(x)&=\ln(x+\sqrt{x^2-1})\,,  \notag \\
\Theta(1-x^2) \cos^{-1}(x)&=-i \ln(x+i \sqrt{1-x^2})\,,  \notag \\
\cos^{-1}(-x)&=\pi-\cos^{-1}(x) \,.  \notag
\end{align}

Thus the functions $G_{>}(x),G_{<}(x)$ can be comprised by the single
function $G(x)=x\sqrt{x^{2}-1}-\ln (x+\sqrt{x^{2}-1})$ where
\begin{equation}
G(x)=\left\{
\begin{array}{ll}
G_{>}(x) & ;\;x>1 \\
iG_{<}(x)=-i[\pi +G_{<}(-x)] & ;\;|x|<1%
\end{array}%
\right. .
\end{equation}

We also note the relation $f(q,\omega)=|F(q,\omega)|$, where
$f(q,\omega)$ is the real function defined in Eq. (\ref{eq:realG})
while $F(q,\omega)$ is the complex function introduced in Eq.
(\ref{eq:complexG}).

\end{document}